\documentclass[12pt,epsf]{article}
\usepackage{amsmath}
\usepackage{amssymb}
\begin{document}
\def\a{\alpha}
\def\b{\beta}
\def\c{\varepsilon}
\def\d{\delta}
\def\e{\epsilon}
\def\f{\phi}
\def\g{\gamma}
\def\h{\theta}
\def\k{\kappa}
\def\l{\lambda}
\def\m{\mu}
\def\n{\nu}
\def\p{\psi}
\def\q{\partial}
\def\r{\rho}
\def\s{\sigma}
\def\t{\tau}
\def\u{\upsilon}
\def\v{\varphi}
\def\w{\omega}
\def\x{\xi}
\def\y{\eta}
\def\z{\zeta}
\def\D{{\mit \Delta}}
\def\G{\Gamma}
\def\H{\Theta}
\def\L{\Lambda}
\def\F{\Phi}
\def\P{\Psi}

\def\S{\Sigma}

\def\o{\over}
\def\beq{\begin{eqnarray}}
\def\eeq{\end{eqnarray}}
\newcommand{\gsim}{ \mathop{}_{\textstyle \sim}^{\textstyle >} }
\newcommand{\lsim}{ \mathop{}_{\textstyle \sim}^{\textstyle <} }
\newcommand{\vev}[1]{ \left\langle {#1} \right\rangle }
\newcommand{\bra}[1]{ \langle {#1} | }
\newcommand{\ket}[1]{ | {#1} \rangle }
\newcommand{\EV}{ {\rm eV} }
\newcommand{\KEV}{ {\rm keV} }
\newcommand{\MEV}{ {\rm MeV} }
\newcommand{\GEV}{ {\rm GeV} }
\newcommand{\TEV}{ {\rm TeV} }
\def\slash#1{\ooalign{\hfil/\hfil\crcr$#1$}}
\def\diag{\mathop{\rm diag}\nolimits}
\def\Spin{\mathop{\rm Spin}}
\def\SO{\mathop{\rm SO}}
\def\O{\mathop{\rm O}}
\def\SU{\mathop{\rm SU}}
\def\U{\mathop{\rm U}}
\def\Sp{\mathop{\rm Sp}}
\def\SL{\mathop{\rm SL}}
\def\tr{\mathop{\rm tr}}

\baselineskip 0.7cm

\begin{titlepage}

\begin{flushright}
UCB-PTH-08/78
\end{flushright}

\vskip 1.35cm
\begin{center}
{\large \bf Emerging AdS from Extremally Rotating NS5-branes}
\vskip 2.2cm

Yu Nakayama

\vskip 0.4cm
{\it Berkeley Center for Theoretical Physics and Department of Physics}
\\

\vskip 3.5cm

\abstract{We investigate the near-horizon limit of extremally rotating NS5-branes. The resulting geometry has $SL(2,\mathbf{R}) \times U(1)^2$ isometry. The asymptotic symmetry group contains a chiral Virasoro algebra, and we obtain two different realizations depending on the boundary conditions we impose.
When one of the two angular momenta vanishes, the symmetry is enhanced to $AdS_3$. The entropy of the boundary theory can be estimated from the Cardy formula and it agrees with the Bekenstein-Hawking entropy of the bulk theory. We can embed the extremally rotating NS5-brane geometry in an exactly solvable string background, which may yield  microscopic understanding of this duality, especially about the mysterious enhancement of the symmetry from $AdS_2$ to $AdS_3$. The construction suggests emerging Virasoro symmetries in the extreme corner of the (1+5) dimensional little string theory.}
\end{center}
\end{titlepage}

\setcounter{page}{2}

\section{Introduction}
The AdS/CFT correspondence, or more generally gauge/gravity correspondence now has become a ubiquitous tool in theoretical physics. It has applications not only to particle physics, but also nuclear (hadron) physics as well as condensed-matter physics. The concept of the gauge/gravity correspondence seems independent of the underlying quantum gravity like string theory, and it was indeed shown long before the advent of the AdS/CFT correspondence for $\mathcal{N}=4$ super Yang-Mills theory \cite{Maldacena:1997re} that the $AdS_3$ space might be holographically dual to a two-dimensional CFT with the Virasoro algebra \cite{Brown:1986nw}.

Quite recently, the authors of \cite{Guica:2008mu} proposed a new holographic duality ``the Kerr/CFT correspondence" based on the similar symmetry argument given  in \cite{Brown:1986nw}. They showed that the four-dimensional extremally rotating Kerr black hole with angular momentum $J=G_NM^2$ is holographically dual to a boundary theory that has a chiral Virasoro algebra with central charge $c = 12J$. In particular, the Bekenstein-Hawking entropy of the extremally rotating Kerr black hole agrees with the entropy of the hypothetical dual CFT by assuming the Cardy formula. Subsequent works on the subject include \cite{Lu:2008jk}\cite{Azeyanagi:2008kb}\cite{Hartman:2008pb}.

It is highly desirable, however, to embed this new holographic duality into string theory. Without such an embedding, it is difficult to recognize what the dual CFT is. If you could not go beyond the kinematical symmetry argument of the correspondence, the dynamical prediction from the correspondence is quite limited. The AdS/CFT correspondence is marvelous simply because it is a duality between well-defined theories in both sides and hence it has a predictive power.

In this paper, we construct a stringy analogue of the Kerr-CFT correspondence based on the extremally rotating NS5-branes. We show that the extremally rotating limit of the geometry has $AdS_2$ isometry with two independent chiral Virasoro algebras acting on the boundary. Furthermore, when one of the two angular momenta vanishes, the symmetry is enhanced to $AdS_3$. A crucial observation is that our background can be realized within the string perturbation theory as an exactly solvable world-sheet CFT. This gives us a possibility to interpret the microscopic origin of this new duality.

Of course, the Kerr black hole {\it is} an approximate solution of string theory compactified down to four dimensions. The point is that in our case, we know the exact background while we do not know such a world-sheet description for the Kerr black hole. The stringy interpretation of the background from the abstract CFT viewpoint will yield microscopic understanding of this duality, especially about the mysterious enhancement of the symmetry from $AdS_2$ to $AdS_3$ as we will see.

The microscopic description of the dual boundary theory is given by the deformation of the little string theory \cite{Aharony:1998ub} realized on NS5-branes. The extremally rotating limit of the deformation suggests emerging Virasoro symmetries in the extreme corner of the moduli space. It is surprising that such an infinite symmetry can appear in the (1+5) dimensional theory. We hope that our exact world-sheet description should shed some light on this duality.


\section{AdS geometry from extremally rotating NS5-brane}
In this section, we study the extremally rotating $N$ NS5-branes background in the supergravity approximation. As we will see, we have an $\alpha'$ exact stringy background corresponding to the supergravity solution, but in this section we focus on the large $N$ limit, where the stringy corrections can be neglected. The supergravity solution was first obtained by the dimensional reduction of the rotating M5-brane background of \cite{Cvetic:1996ek} in \cite{Sfetsos:1999pq}. One can also obtain the solution from a certain null-Melvin twist of the non-rotating black NS5-branes solution \cite{Maldacena:1997cg} (see section 4 for more details).

We begin with the rotating NS5-branes solution in the so-called ``field theory limit" \cite{Sfetsos:1999pq}:\footnote{We use $\alpha'=1$ notation.}
\begin{align}
\frac{1}{N} ds^2 &= -\left(1-\frac{1}{\Delta_0}\right) d\tilde{t}^2 + dy_1^2 + \cdots +dy_5^2 +\frac{d\tilde{\rho}^2}{\tilde{\rho}^2 + a^2_1a^2_2/\tilde{\rho}^2 + a_1^2 + a_2^2-1} \cr
& + d\theta^2 + \frac{1}{\Delta_0}\left((\tilde{\rho}^2+a_1^2)\sin^2\theta d\tilde{\phi}_1^2+ (\tilde{\rho}^2 + a_2^2) \cos^2\theta d\tilde{\phi}_2^2\right) \cr
&-\frac{2}{\Delta_0} d\tilde{t} (a_1\sin^2\theta d\tilde{\phi}_1 + a_2\cos^2\theta d\tilde{\phi}_2) \ , \label{metr}
\end{align}
where $\Delta_0 = \tilde{\rho}^2 + a_1^2\cos^2\theta + a_2^2 \sin^2\theta$
with the Kalb-Ramond two-form
\begin{eqnarray}
\frac{1}{N} B = 2\frac{1}{\Delta_0} \left(-(\tilde{\rho}^2 + a_1^2) \cos^2\theta d\tilde{\phi}_1 \wedge d\tilde{\phi}_2 + a_2\sin^2\theta d\tilde{t} \wedge d\tilde{\phi}_1 + a_1\cos^2\theta d\tilde{t} \wedge d\tilde{\phi}_2 \right) \ 
\end{eqnarray}
and the dilaton gradient
\begin{eqnarray}
e^{2\Phi} = \frac{N}{\mu\Delta_0} \ .
\end{eqnarray}
The geometry \eqref{metr} is not asymptotically flat, but rather a linear dilaton theory with a linear dilaton slope $Q = \frac{1}{\sqrt{N}}$. However, it is possible to embed the solution \eqref{metr} into an asymptotically flat background (see \cite{Sfetsos:1999pq}). After this embedding, $\mu$ is inversely proportional to the asymptotic string coupling constant $g_s$ and hence the Newton constant $G_N$ of the asymptotically flat background: $\mu \propto \frac{1}{g_s^2} \propto \frac{1}{G_N}$ \ .

Since the metric \eqref{metr} is not asymptotically flat, it is difficult to read the thermodynamic quantities from the solution directly, but with the embedding to the asymptotic flat space mentioned above, one can read the thermodynamic quantities by using conventional methods in general relativity. These thermodynamic quantities: angular momenta $J_{i}$ $(i=1,2)$, angular velocities $\Omega_i$, Bekenstein-Hawking entropy $S$, and Hawking temperature $T_H$, can be computed as 
\begin{align}
J_i &= \frac{\Omega_3 V_5}{8\pi} \mu a_i \sqrt{N} \cr
\Omega_i &= \frac{a_i}{(\rho_H^2 + a_i^2)\sqrt{N}} \cr
S &= \frac{\Omega_3 V_5}{4} \mu \rho_H \sqrt{N} \cr
T_H &= \frac{\rho_H^4 - a_1^2a_2^2}{ 2\pi \rho_H^3 \sqrt{N}} \ , \label{thermo} 
\end{align}
where $\Omega_3 = 2\pi^2$ is the volume of the 3-sphere and $V_5$ is that of the tangential flat 5-dimensional space spanned by $(y_1, \cdots, y_5)$. The outer horizon position $\rho_H$ is given by
\begin{eqnarray}
\rho_H^2 = \frac{1}{2} \left(1 - a_1^2 -a_2^2 + \sqrt{(1-a_1^2-a_2^2)^2 - 4a_1^2a_2^2}\right) \ . \label{horiz}
\end{eqnarray}
The inner horizon is located at the position where you replace the sign in front of the square-root in \eqref{horiz}.

We are interested in the extremally rotating limit $a_1 + a_2 = 1$, in which the outer horizon and the inner horizon coincide. In this limit, the metric \eqref{metr} can be rewritten as
\begin{align}
\frac{1}{N} ds^2 &= -\frac{(\tilde{\rho}^2-a_1a_2)^2}{(\tilde{\rho}^2+a_1^2)(\tilde{\rho}^2+a_2^2)} d\tilde{t}^2 + dy_1^2 + \cdots dy_5^2 +\frac{d\tilde{\rho}^2}{(\tilde{\rho}-\frac{a_1a_2}{\tilde{\rho}})^2} \cr
& + d\theta^2 + \frac{1}{\Delta_0}(\tilde{\rho}^2+a_1^2) \sin^2\theta \left(d\tilde{\phi}_1 - \frac{a_1d\tilde{t}}{\tilde{\rho}^2+a_1^2}\right)^2 \cr 
&+ \frac{1}{\Delta_0}(\tilde{\rho}^2+a_2^2) \cos^2\theta \left(d\tilde{\phi}_2 - \frac{a_2d\tilde{t}}{\tilde{\rho}^2+a_2^2}\right)^2 \ .
\end{align}
The horizon is now located at $\tilde{\rho}^2 = a_1a_2$. In this extremally rotating limit, the temperature is zero: $T_H =0$ while the Bekenstein-Hawking entropy is finite $S= \frac{\Omega_3 V_5}{4} \mu \sqrt{a_1a_2 N}$.

Following \cite{Bardeen:1999px}, we investigate the near-horizon limit of the geometry by setting 
\begin{eqnarray}
\tilde{t} = \frac{t}{2\lambda} \ , \ \ \tilde{\rho}^2 = a_1 a_2 +  \sqrt{a_1a_2}\lambda y \ , \ \ 
\end{eqnarray}
and taking $\lambda \to 0$. The resulting metric is given by
\begin{align}
\frac{ds^2}{N} &= \frac{1}{4}\left(-y^2 dt^2 + \frac{dy^2}{y^2}\right) + dy_1^2 + \cdots +dy_5^2 +d\theta^2 \cr
&+ \frac{a_1\sin^2 \theta}{a_1a_2 + a_1^2\cos^2\theta + a_2^2 \sin^2\theta} \left(d{\phi}_1 +  \sqrt{\frac{a_2}{a_1}} \frac{y}{2} dt \right)^2 \cr 
&+ \frac{a_2\cos^2 \theta}{a_1a_2 + a_1^2\cos^2\theta + a_2^2 \sin^2\theta} \left(d{\phi}_2 + \sqrt{\frac{a_1}{a_2}} \frac{y}{2} dt \right)^2  \ ,
\end{align}
where ${\phi_i} = \tilde{\phi}_i - \tilde{t}$.
The geometry is (deformed) 3-sphere bundle over the $AdS_2$ space.  The isometry group is $SL(2,\mathbf{R}) \times U(1)^2$ (and the Euclidean group for $\mathbf{R}^5$). 
We also have the background Kalb-Ramond field
\begin{align}
\frac{B}{N} &= -2 \frac{a_1\cos^2\theta}{a_1a_2 + a_1^2 \cos^2\theta + a_2^2\sin^2\theta} d{\phi}_1 \wedge d{\phi}_2 \cr
 &- \frac{\sqrt{a_1a_2} y \cos^2\theta}{a_1a_2 + a_1^2 \cos^2\theta + a_2^2 \sin^2\theta} dt \wedge d\phi_2 - \frac{\sqrt{a_1a_2}y \sin^2\theta}{a_1a_2 + a_1^2 \cos^2\theta + a_2^2 \sin^2\theta} dt \wedge d\phi_1  \ 
\end{align}
together with the dilaton gradient
\begin{eqnarray}
e^{2\Phi} = \frac{N}{\mu (a_1a_2  + a_1^2\cos^2\theta + a_2^2\sin^2\theta)} \ .
\end{eqnarray}

In particular, in the special case of equal angular momenta: $a_1 = a_2 = \frac{1}{2}$, the $\theta$ dependence in $\Delta_0$ disappears, and we have a round $S_3$ with the simple metric
\begin{eqnarray}
\frac{ds^2}{N} &= \frac{1}{4}\left(-y^2 dt^2 + \frac{dy^2}{y^2}\right) +  dy_1^2 + \cdots +dy_5^2+d\theta^2 \cr &+ \sin^2\theta \left(d{\phi}_1 + \frac{y}{2} dt\right)^2 + \cos^2\theta \left(d{\phi}_2 + \frac{y}{2} dt\right)^2 \ .
\end{eqnarray}
The Kalb-Ramond field is given by
\begin{eqnarray}
\frac{1}{N} B = -2 \cos^2\theta d{\phi}_1 \wedge d{\phi}_2 - y \cos^2\theta dt \wedge d{\phi}_2 - y \sin^2\theta dt \wedge d{\phi}_1 \ .
\end{eqnarray}
The dilaton is everywhere constant in this particular case.

It is easy to extend the Poincare-like metric to a global patch (c.f. \cite{Bardeen:1999px}) by defining
\begin{align}
y &= r + \sqrt{1+{r}^2} \cos \tau \cr
t &= \frac{\sqrt{1+{r}^2} \sin\tau}{{r} + \sqrt{1+{r}^2} \cos \tau} \ .
\end{align}
The near-horizon metric now reads 
\begin{align}
\frac{ds^2}{N} &= \frac{1}{4}\left(-(1+{r}^2)d\tau^2 + \frac{d{r}^2}{1+{r}^2}\right) + dy_1^2 \cdots + dy_5^2 + d\theta^2 \cr &+ \frac{a_1\sin^2 \theta}{a_1a_2 + a_1^2\cos^2\theta + a_2^2 \sin^2\theta} \left(d{\phi}_1 + \sqrt{\frac{a_2}{a_1}} \frac{r}{2} d\tau \right)^2 \cr 
&+ \frac{a_2\cos^2 \theta}{a_1a_2 + a_1^2\cos^2\theta + a_2^2 \sin^2\theta} \left(d{\phi}_2 + \sqrt{\frac{a_1}{a_2}}\frac{r}{2} d\tau \right)^2  \ ,
\end{align}
where we have shifted $d\phi_i$ by an amount proportional to $d \log \left(\frac{1+\sqrt{1+r^2} \sin \tau}{\cos\tau + r\sin\tau} \right)$.

The symmetry will be enhanced if we set one of the angular momenta (say $a_2$) zero. To see this, we turn back to the original metric \eqref{metr}.
By setting $a_2 = 0$ and defining a new coordinate as $\tilde{\rho} = \lambda r$, $\tilde{t} = \frac{t}{\lambda}$, and $\tilde{\phi}_2 = \frac{\phi_2}{\lambda}$, the metric becomes
\begin{eqnarray}
\frac{ds^2}{N} = -r^2 dt^2 + \frac{dr^2}{r^2} +  dy_1^2 + \cdots +dy_5^2+ d\theta^2 +\tan^2\theta d{\phi}_1^2 + r^2 d\phi_2^2 \ . \label{ax}
\end{eqnarray}
The geometry is given by the direct product of $AdS_3$ (spanned by $r,t$ and $\phi_2$) and disk (spanned by $\theta$ and $\phi_1$).\footnote{The literal limit demands an infinitesimal periodicity for $\phi_2$, but there is no obstacle in decompactifying this direction.} The obvious isometry is $SL(2,\mathbf{R}) \times SL(2,\mathbf{R}) \times U(1)$ (and the Euclidean group acting on $\mathbf{R}^5$).\footnote{The $AdS_3$ part naturally gives rise to boundary Virasoro algebras from the analysis of \cite{Brown:1986nw}.} It seems that the exact CFT description is possible by $SL(2,\mathbf{R})$ WZW model and $SU(2)/U(1)$ coset model as we will see in section 4. The Kalb-Ramond field becomes
\begin{eqnarray}
\frac{1}{N}B = -2r^2 dt \wedge d\phi_2 \ ,
\end{eqnarray}
and the dilaton gradient is 
\begin{eqnarray}
e^{2\Phi} = \frac{N}{\mu \cos^2\theta} \ .
\end{eqnarray}
Note that the Kalb-Ramond field only appears in the $AdS_3$ part and the non-trivial dilaton gradient is in the disk part as is consistent with the exact CFT construction.

Similarly, for $a_1 = 0$, we have
\begin{align}
\frac{ds^2}{N} &= -r^2 dt^2 +\frac{dr^2}{r^2} +  dy_1^2 + \cdots +dy_5^2 + d\theta^2 + \cot^2\theta d{\phi}_2^2 + r^2 d\phi_1^2 \cr 
\frac{1}{N}B &= -2r^2 dt \wedge d\phi_1 \cr
e^{2\Phi} &= \frac{N}{\mu \sin^2\theta} \ . \label{ve}
\end{align}
The geometry is related to $a_2=0$ case by T-duality in $\phi_2$ direction (up to exchange of $\phi_2$ and $\phi_1$). In the exact coset description, the axial gauging of $SU(2)/U(1)$ gives \eqref{ax} while the vector gauging of $SU(2)/U(1)$ gives \eqref{ve}.

\section{Central charge, Temperature and Entropy}
We can employ the method developed in \cite{Guica:2008mu} to compute the central charge of the dual (chiral) CFT. It turns out that the boundary theory corresponding to the extremally rotating NS5-branes admits two copies of chiral Virasoro algebra.

We assume the similar ansatz for the boundary condition proposed in \cite{Guica:2008mu}\cite{Azeyanagi:2008kb}: for the fluctuation of the metric $h_{\mu\nu}$, we require either the boundary condition (1)
\begin{align} 
\left(
  \begin{array}{ccccc}
   h_{\tau\tau} = O(1)& h_{\tau r}=O(r^{-2})& h_{\tau \theta}= O(r^{-1}) &h_{t\phi_1} =O(1) & h_{\tau \phi_2}=O(r) \\
   h_{r\tau}=h_{\tau r}  & h_{rr}=O(r^{-3}) & h_{r \theta} = O(r^{-2}) & h_{r\phi_1} = O(r^{-1}) & h_{r\phi_2} = O(r^{-1})  \\
 h_{\theta \tau} = h_{\tau \theta} & h_{\theta r} = h_{r\theta} & h_{\theta\theta} = O(r^{-1}) & h_{\theta \phi_1} = O^(r^{-1}) & h_{\theta \phi_2} = O(r^{-1}) \\
h_{\phi_1 \tau} = h_{\tau \phi_1} & h_{\phi_1 r} = h_{r\phi_1} & h_{\phi_1 \theta} = h_{\theta \phi_1} & h_{\phi_1\phi_1} = O(1) & h_{\phi_1 \phi_2} = O(1) \\
h_{\phi_2 \tau } = h_{\tau \phi_2} & h_{\phi_2 r} = h_{r\phi_2} & h_{\phi_2 \theta} = h_{\theta \phi_2} & h_{\phi_2 \phi_1} = h_{\phi_1\phi_2} & h_{\phi_2\phi_2} = O(r^{-1}) \\
  \end{array}
\right) 
\end{align}
or the boundary condition (2)
\begin{align} 
\left(
  \begin{array}{ccccc}
    h_{\tau\tau} = O(1)& h_{\tau r}=O(r^{-2})& h_{\tau \theta}= O(r^{-1}) &h_{\tau \phi_1} =O(r) & h_{\tau \phi_2}=O(1) \\
   h_{r\tau}=h_{\tau r}  & h_{rr}=O(r^{-3}) & h_{r \theta} = O(r^{-2}) & h_{r\phi_1} = O(r^{-1}) & h_{r\phi_2} = O(r^{-1})  \\
 h_{\theta \tau} = h_{\tau \theta} & h_{\theta r} = h_{r\theta} & h_{\theta\theta} = O(r^{-1}) & h_{\theta \phi_1} = O^(r^{-1}) & h_{\theta \phi_2} = O(r^{-1}) \\
h_{\phi_1 \tau } = h_{\tau \phi_1} & h_{\phi_1 r} = h_{r\phi_1} & h_{\phi_1 \theta} = h_{\theta \phi_1} & h_{\phi_1\phi_1} = O(r^{-1}) & h_{\phi_1 \phi_2} = O(1) \\
h_{\phi_2 \tau} = h_{\tau \phi_2} & h_{\phi_2 r} = h_{r\phi_2} & h_{\phi_2 \theta} = h_{\theta \phi_2} & h_{\phi_2 \phi_1} = h_{\phi_1\phi_2} & h_{\phi_2\phi_2} = O(1) \\
  \end{array}
\right) 
\end{align}
Correspondingly, the asymptotic symmetry group is generated by the following vector fields
\begin{align}
\zeta^{(1)}_{(n)} &= -e^{-in\phi_1} \frac{\partial}{\partial \phi_1} - i n r e^{-in \phi_1} \frac{\partial}{\partial r} \cr
 \zeta^{(2)}_{(n)} & =  -e^{-in\phi_2} \frac{\partial}{\partial \phi_2} - i n r e^{-in \phi_2} \frac{\partial}{\partial r} \ .
\end{align}
We see that the commutators of these vector fields generate two copies of chiral Virasoro algebra with zero central charge:
\begin{eqnarray}
i[\zeta^{(i)}_{(m)},\zeta^{(j)}_{(n)}] = \delta^{ij} (m-n) \zeta_{(n+m)}^{(i)} \  , 
\end{eqnarray}
However, we note that the only one of the Virasoro generators can be realized at a time as an asymptotic symmetry group due to the different boundary conditions imposed as above \cite{Azeyanagi:2008kb}.\footnote{We would like to thank the referee for pointing out this fact and referring to the literature.}

The corresponding charges are defined by\footnote{Strictly speaking, the diffeomorphism also acts on the Kalb-Ramond field and dilaton, and we have corresponding charges. 
Recently, it has been shown that the central charge contribution from the gauge field vanishes in the Kerr-Newmann black hole \cite{Hartman:2008pb}, and we expect the similar situation here. The agreement of the entropy without the correction from the Kalb-Ramond field supports this argument.} 
\begin{eqnarray}
Q_{\zeta} = \frac{\mu}{8\pi N} \int_{\partial \Sigma} k_{\zeta}[h,g] \ ,
\end{eqnarray}
where $\partial \Sigma$ is a spatial slice, and the 3-form $k_{\zeta}$ is given by
\begin{align}
k_{\zeta}[h,g] =& \frac{1}{2} \left[\zeta_\nu D_\mu h - \zeta_\nu D_\sigma h_{\mu}^{\ \sigma} + \zeta_{\sigma} D_\nu h_{\mu}^{\ \sigma} + \frac{1}{2} h D_{\nu} \zeta_\mu - h_{\nu}^{\ \sigma} D_\sigma\zeta_\mu + \frac{1}{2} h^{\sigma}_{\ \nu} (D_\mu \zeta_\sigma + D_{\sigma} \zeta_{\mu})\right] \cr
 &*(dx^\mu\wedge dx^\nu) \ ,
\end{align}
where $*$ denotes the Hodge dual in the five-dimension ($\tau,r,\theta, \phi_1,\phi_2$) multiplied by $(a_1a_2 + a_1^2 \cos^2\theta + a_2^2\sin^2\theta)$ from the string measure.

The Dirac bracket yields the central term in the Virasoro algebra \cite{Barnich:2001jy}:
\begin{eqnarray}
\frac{\mu}{8\pi N} \int_{\partial \Sigma} k_{\zeta_{(m)}^{(i)}}[\mathcal{L}_{\zeta_{(n)}^{(i)}}g,g] = -\frac{i}{12} (m^3 + xm) \delta_{m+n} c^{(i)} \ , \label{vir}
\end{eqnarray}
where $\mathcal{L}_\zeta g_{\mu\nu} = \zeta^\sigma \partial_\sigma g_{\mu\nu} + g_{\sigma\nu} \partial_\mu \zeta^{\sigma} + g_{\mu\sigma} \partial_\nu \zeta^{\sigma}$ is the Lie derivative of the metric with respect to the vector field $\zeta$. $c^{(i)}$ gives the central charge of the (chiral) Virasoro algebra. The coefficient $x$ is unimportant because one can always absorb it by the shift of the Virasoro zero mode $L_0 \to L_0 + \delta_0$ with a certain constant $\delta_0$. The choice of $x=-1$ is conventional.

The relevant Lie derivatives can be computed as
\begin{align}
\mathcal{L}_{\zeta^{(1)}_{(n)}} g_{\tau\tau} &= 0\cr
\mathcal{L}_{\zeta^{(1)}_{(n)}} g_{r\phi_1} &= -\frac{1}{4}\frac{Nr}{1+r^2}n^2 e^{-in\phi_1} \cr
\mathcal{L}_{\zeta^{(1)}_{(n)}} g_{\phi_1\phi_1} &=\frac{2Na_1\sin^2\theta}{a_1a_2 + a_1^2 \cos^2\theta + a_2^2\sin^2\theta} in e^{-in\phi_1} \cr
\mathcal{L}_{\zeta^{(1)}_{(n)}} g_{rr} &= -\frac{N}{2(1+r^2)^2} ine^{-in\phi_1}
\end{align}
and similarly for $\zeta^{(2)}_{(n)}$. Substituting them into \eqref{vir}, we can compute the central charges as
\begin{eqnarray}
c^{(1)} = \frac{3}{2}\mu \pi a_2 \sqrt{N} \ ,   \ \ c^{(2)} = \frac{3}{2}\mu \pi a_1 \sqrt{N} \ . 
\end{eqnarray}
Note that the two chiral Virasoro algebras have different central charges (see a similar situation in the Myers-Perry black hole studied in \cite{Lu:2008jk}).

Let us move on to the computation of the temperature of the boundary theories. Again we will follow the strategy proposed in \cite{Hartman:2008pb}.
We define the temperature of the extremally rotating NS5-branes from the Frolov-Thorne vacuum \cite{Frolov:1989jh}. In this vacuum, we expand the quantum field in the rotating NS-branes background by the eigenstates\footnote{The factor $\sqrt{N}$ in front of the energy $\omega$ is due to the rescaling of $\tilde{t} \to \sqrt{N} \tilde{t}$ compared with the asymptotic time.}
\begin{eqnarray}
 e^{-i\sqrt{N}\omega  \tilde{t} + i m_1 \tilde{\phi}_1 + im_2 \tilde{\phi}_2} = e^{-in_R t + i n_{\phi_1} \phi_1 + in_{\phi_2} \phi_2} \ , \label{energ}
\end{eqnarray}
or
\begin{eqnarray}
n_R =\frac{\sqrt{N}\omega}{2\lambda} -\frac{m_1}{2\lambda} -\frac{m_2}{2\lambda}  \ , \ \ n_{\phi_1} = m_1 \ , \ \ n_{\phi_2} = m_2 \ .
\end{eqnarray}

With these near-horizon variables, the Boltzmann factor can be rewritten as
\begin{eqnarray}
e^{-\frac{\omega - \Omega_1 m_1 - \Omega_2 m_2}{T_H}} = e^{-\frac{n_R}{T_R} - \frac{n_{\phi_1}}{T_{\phi_1}} -\frac{n_{\phi_2}}{T_{\phi_2}}} \ .
\end{eqnarray}
Explicitly
\begin{eqnarray}
T_R = \frac{T_H}{2\lambda} \ , \ \ T_{\phi_1} = \frac{\sqrt{N}T_H}{1-\sqrt{N}\Omega_1} \ , \ \ T_{\phi_2} = \frac{\sqrt{N}T_H}{1-\sqrt{N}\Omega_2} \ .
\end{eqnarray}
By taking the extremally rotating limit $a_1 + a_2 = 1$ and by using \eqref{thermo}, we obtain
\begin{eqnarray}
T_R = 0 \ , \ \ T_{\phi_1} = \frac{a_1}{\pi\sqrt{a_1a_2}} \ , \ \ T_{\phi_2} = \frac{a_2}{\pi\sqrt{a_1a_2}} \ . 
\end{eqnarray}

Finally, we can estimate the entropy of the boundary theory by using the Cardy formula. According to the Cardy formula, the entropy of a unitary compact CFT (about which we simply assume) is given by
\begin{equation}
S = \frac{\pi^2}{3} c T \ .
\end{equation}
In our case, although the two Virasoro algebras have different central charges as well as temperature, the both entropy coincide
\begin{align}
 S &= S_1 = \frac{\pi^2}{3} c^{(1)}T_{\phi_1} \cr
   &= S_2 = \frac{\pi^2}{3} c^{(2)}T_{\phi_2} \cr
   &= \frac{\pi^2}{2} \mu \sqrt{a_1a_2 N} \ . 
\end{align}
We note that in contrast to the case \cite{Guica:2008mu}, one can parametrically meet the sufficient condition of the Cardy formula $c\gg T$ in one of the Virasoro algebras (but not both) by tuning $a_i \to 0$.
This expression agrees with the Bekenstein-Hawking entropy density (i.e. up to a volume factor $V_5$) computed from the bulk geometry \eqref{thermo}. The field theory entropy is an extensive quantity, so it is natural that the boundary computation discussed in this section, which is valid point-wise in tangential flat 5 dimension, expects the volume factor $V_5$.

\section{Exact Solution in String theory}
The supergravity solution \eqref{metr} has an exact stringy description based on $SL(2,\mathbf{R})_N/U(1)$ coset model and $SU(2)_N$ WZW model. It is easy to see the geometric origin of these exact CFTs. The supergravity solution  \eqref{metr} is asymptotically linear dilaton theory with the slope $Q = \frac{1}{\sqrt{N}}$, which suggests the two-dimensional black hole geometry given by $SL(2,\mathbf{R})_N/U(1)$ coset model with the same linear dilaton slope. The $S_3$ part of the metric together with the Kalb-Ramond field is nothing but the supergravity approximation to the $SU(2)_N$ WZW model.

The central charge of the system remains critical due to the linear dilaton contributions
\begin{eqnarray}
c = c_{SL(2,\mathbf{R})_N/U(1)} + c_{SU(2)_N} = \left(3 + \frac{6}{N} \right) + \left(\frac{9}{2}-\frac{6}{N} \right) = \frac{15}{2} \ .
\end{eqnarray}
The rest of the central charge is provided by the flat Euclidean space $\mathbf{R}^{5}$ for ($y_1, \cdots,y_5$).

The (Euclidean version of the) supergravity solution corresponding to the field theory limit of the rotating NS5-branes can be obtained by an $O(3,3)$ transformation of the non-rotating solution given by the $ SL(2,\mathbf{R})_N/U(1) \times SU(2)_N$ exactly solvable CFT background as noted in \cite{Sfetsos:1999pq}. In this section, we give physical intuitive understanding of this $O(3,3)$ transformation as a certain null-Melvin twist in a step-by-step manner.\footnote{Related rotating NS5-brane solutions have been studied in \cite{Itzhaki:2005zr}\cite{Nakayama:2005}.}

To construct the non-extremally rotating NS5-brane solution in \cite{Sfetsos:1999pq}, we begin with the non-extremal (Lorentzian) 2D black hole (i.e. $SL(2,\mathbf{R})_N/U(1)$ coset) solution with an additional $U(1)$ circle\footnote{We focus on the rotation in one directions (i.e. $a_2 = 0$). A rather trivial generalization of the three-dimensional boost gives non-zero angular momenta in both rotating directions.}:
\begin{eqnarray}
ds^2 = N \left(d\hat{\rho}^2 + d\hat{x}^2 - \tanh^2\rho d\hat{t}^2 \right) \ . \label{lbh}
\end{eqnarray}
We perform the Lorentz boost
\begin{align}
\frac{1}{Q} \hat{x} &= \frac{1}{R} {x} + \omega \tilde{t} \cr
\frac{1}{Q} \hat{t} &= \frac{1}{R} \tilde{t} + \omega {x} \ , \label{boost}
\end{align}
where $\omega$ is the boost parameter and the radius $R$ of the $U(1)$ is given by
(recall $Q^2 = \frac{1}{N})$
\begin{eqnarray}
\frac{1}{R^2} \equiv w^2 + \frac{1}{Q^2} \ . \label{rad}
\end{eqnarray}
The metric becomes
\begin{align}
ds^2 &= N\left(d\hat{\rho}^2 + [\omega^2Q^2(1-\tanh^2\hat{\rho}) +1]dx^2 + 2\omega Q\sqrt{1+\omega^2Q^2} (1-\tanh^2\hat{\rho}) d\tilde{t}dx \right. \cr & \left. + [-\tanh^2\hat{\rho} + \omega^2Q^2(1-\tanh^2\hat{\rho}) ]d\tilde{t}^2 \right) \ .
\end{align}

$SU(2)_N$ WZW model can be reconstructed from the $\mathbf{Z}_k$ orbifolding of $U(1)$ part given by $ \hat{\phi}_1 \equiv x$ and the $SU(2)_N/U(1)$ coset model. The coset model part can be represented by the metric
\begin{eqnarray}
ds^2 = N(d\theta^2 + \tan^2\theta d\hat{\phi}^2_2) \ .
\end{eqnarray}
The $\mathbf{Z}_N$ orbifolding (so-called ``GSO projection") is performed by the identification $(\hat{\phi}_1,\hat{\phi}_2) \sim (\hat{\phi}_1 + \frac{2\pi}{N}, \hat{\phi}_2 + \frac{2\pi}{N})$. We can undo this orbifolding by defining a new coordinate
\begin{align}
\hat{\phi}_1 &= \tilde{\phi}_1 + \frac{\tilde{\phi_2}}{N} \cr
\hat{\phi}_2 &= \frac{\tilde{\phi}_2}{N} \ .
\end{align}
In this coordinate, the metric becomes
\begin{align}
ds^2 &= Nd\hat{\rho}^2 + Nd\theta^2 + \frac{1}{N}[\tan^2\theta + 1 + \omega^2Q^2(1-\tanh^2\hat{\rho})] d\tilde{\phi}^2_2 + N[\omega^2Q^2(1-\tanh^2\hat{\rho}) + 1] d\tilde{\phi}_1^2 \cr &+ 2N[\omega^2Q^2(1-\tanh^2\hat{\rho}) + 1] d\tilde{\phi}_1 d\tilde{\phi}_2 + N[-\tanh^2\hat{\rho} + \omega^2Q^2(1-\tanh^2\hat{\rho})] d\tilde{t}^2 \cr  &+ 2N\omega Q\sqrt{1+\omega^2Q^2} (1-\tanh^2\hat{\rho}) d\tilde{t}(d\tilde{\phi}_1 + \frac{d\tilde{\phi}_2}{N}) \ , 
\end{align}

Now we take the T-duality along the $\tilde{\phi}_2$ direction. The resulting geometry reads
\begin{align}
g_{22} &= N\frac{\cosh^2\hat{\rho}}{\cosh^2 \hat{\rho} + \omega^2Q^2 \cos^2\theta}\cos^2\theta \cr
g_{11} &= N\frac{\cosh^2\hat{\rho} + \omega^2 Q^2}{\cosh^2 \hat{\rho} + \omega^2Q^2 \cos^2\theta}\sin^2\theta \cr
g_{\hat{t}\hat{t}} &=N\left( -1 + \frac{1+\omega^2Q^2}{\cosh^2 \hat{\rho} + \omega^2Q^2 \cos^2\theta}\right) \cr
g_{\hat{t}1} &= N\omega Q \sqrt{1+ \omega^2Q^2} \frac{1}{\cosh^2 \hat{\rho} + \omega^2Q^2 \cos^2\theta}\sin^2\theta \cr
B_{\hat{t}2} &= N\omega Q \sqrt{1+ \omega^2Q^2} \frac{1}{\cosh^2 \hat{\rho} + \omega^2Q^2 \cos^2\theta}\cos^2\theta \cr
B_{21} &= N\frac{\cosh^2\hat{\rho} + \omega^2 Q^2}{\cosh^2 \hat{\rho} + \omega^2Q^2 \cos^2\theta}\cos^2\theta \ , \label{local}
\end{align}
which is equivalent to the non-extremally rotating NS5-brane solution presented in  \eqref{metr} (for $a_2 = 0$). To compare \eqref{local} with \eqref{metr}, we define $\omega^2 Q^2 = \frac{a_1^2}{1-a_1^2}$ and extend the coordinate $\hat{\rho}$ to $\tilde{\rho}$ by $(1-a_1^2)\cosh^2\hat{\rho} = \tilde{\rho}^2$. Note that this solution has a finite horizon in the extended coordinate, and $\hat{\rho}$ coordinate covers the outside of it.

In order to take the extremally rotating limit $a_1 +a_2 =1$, it is clear that we have to do infinite boost $\omega \to \infty$. This is roughly the definition of ``null"-Melvin twist. With this regard, as studied in section 2, it is worth noticing that the supergravity solution with only one angular momentum (i.e. $a_2 = 0$ or $a_1=0$) describes the $AdS_3$ space. We can also construct the exact background corresponding to \eqref{ax} or \eqref{ve}. The former is given by $SL(2,\mathbf{R})_N \times SU(2)_N^{(A)}/U(1)$ coset, where $(A)$ means the axial coset while the latter is given by $SL(2,\mathbf{R})_N \times SU(2)_N^{(V)}/U(1)$, where $(V)$ means the vector coset. The axial coset and vector coset are related by the T-duality, so as a CFT, the background is essentially the same.

In the exact CFT approach, the enhancement of $AdS_2$ symmetry to $AdS_3$ symmetry is not unexpected because the $SL(2,\mathbf{R})_N/U(1)$ coset theory already knows the structure of the parent $SL(2,\mathbf{R})_N$ WZW model. The stringy understanding of this symmetry enhancement should shed some light on the rather mysterious phenomena in the supergravity limit.\footnote{This was first observed in \cite{Bardeen:1999px} in the near-horizon limit of the Myers-Perry black hole.}  See also 
\cite{Israel:2005ek} for a similar construction of $AdS_3$ space (or Lorentzian $SL(2,\mathbf{R})$ WZW model) from the singular deformation of the $SL(2,\mathbf{R})/U(1)$ times $U(1)$ theory. See also \cite{Compere:2008cw} for constructing of rotating black holes from $SL(2,\mathbf{R})$ WZW model.

\section{Comment on Microstate from Little String Theory}
We would like to conclude this paper by proposing the microscopic dual theory interpretations of the duality. The effective theory living on stacks of NS5-branes is known as little string theory \cite{Aharony:1998ub}, and it is natural that our extremally rotating limit corresponds to a particular deformation of the little string theory.

Indeed, the rotating effects of the (extremal) NS5-brane were proposed in \cite{Itzhaki:2005zr}, where the authors argued that the deformation corresponds to a supersymmetry breaking mass term with non-zero time-dependent VEV for the moduli fields $\langle A \rangle \sim e^{-i\omega t}$. The oscillating behavior of the moduli fields is captured by the rotation of the NS5-branes.\footnote{The corresponding supergravity solution is singular, but the singularity is resolved by the $\mathcal{N}=2$ Liouville potential (``winding tachyon condensation") in the exact world-sheet CFT analysis \cite{Nakayama:2005}.}

In our example, we have to incorporate the effect of finite temperature as well before the rotation. This is because our solution is obtained by a null-Melvin twist of non-extremal  NS5-branes, in contrast to the one studied in \cite{Itzhaki:2005zr} that was obtained by the twist of extremal (supersymmetric) NS5-branes. The supergravity solution is not singular in our case. The oscillating energy in the boundary theory given by $\omega$ is infinite here, and as a consequence, the radius of the $U(1)$ is infinitesimal (see \eqref{rad}). This singular behavior demands emergence of new degrees of freedom, and they would explain the emergence of the Virasoro symmetries.

It is quite remarkable that the dual little string theory in this particular corner of the moduli space possesses Virasoro symmetries as a $1+5$ dimensional theory. Note that the Virasoro symmetries are rather internal symmetries than the space symmetry of the little string theory. This is because the Virasoro symmetries here commute with all the Euclidean group acting on $\mathbf{R}^5$. Presumably, the little string theory shows a condensation of string degrees of freedom, which might explain the point-wise Virasoro symmetries we have discovered in this paper.

Finally, our solution may have some relations to supergravity solutions with non-relativistic conformal symmetry  \cite{Son:2008ye}. The non-relativistic conformal backgrounds are also obtained from a null-Melvin twist of the non-rotating black membrane solutions \cite{Herzog:2008wg}\cite{Maldacena:2008wh}\cite{Adams:2008wt}\cite{Schvellinger:2008bf}\cite{Mazzucato:2008tr}. The appearance of the mixed term in the metric $dt dx$ is reminiscent of such a construction, and our solution shows a formal similarity to this.

\section*{Acknowledgements}
The author would like to thank S.~Rey and Y.~Sugawara for stimulating discussions on the rotating NS5-branes back in 2005.
A part of the arguments in section 4 is based on the collaboration with them.
 He also acknowledges T.~Nishioka for pointing out a factor error in the first version of the manuscript.
The research of Y.~N. is supported in part by NSF grant PHY-0555662 and the UC Berkeley Center for Theoretical Physics.

\end{document}